\documentclass[12pt]{article}
\def\today{October 24, 2007} 
\usepackage{amsmath,amssymb}
\usepackage{hyperref}
\newtheorem{thm}{Theorem}[section]
\newtheorem{co}[thm]{Corollary}
\newtheorem{lem}[thm]{Lemma}
      
\newtheorem{assumption}[thm]{Assumption}

\newtheorem{pr}[thm]{Proposition}

\newtheorem{problem}[thm]{Problem}
\newenvironment{prob}{\begin{problem}\rm}{\end{problem}}

\newtheorem{assu1}[thm]{Assumption}

\newtheorem{definition}[thm]{Definition}
\newenvironment{de}{\begin{definition}\rm}{\end{definition}}
\newtheorem{example}[thm]{Example}
\newenvironment{exmp}{\begin{example}\rm}{\end{example}}
\newtheorem{remark}[thm]{Remark}
\newenvironment{rem}{\begin{remark}\rm}{\end{remark}}
\newtheorem{protocol}[thm]{Protocol}
\newenvironment{pro}{\begin{protocol}\rm}{\end{protocol}}

\newcommand{\F}{\mathbb{F}}

\newcommand{\N}{\mathbb{N}}
\newcommand{\Z}{\mathbb{Z}}
\newcommand{\mat}{\mathrm{Mat}}

\newcommand{\lcm}{\mathrm{lcm}\,}

\newcommand{\ord}{\mathrm{ord}}
\newcommand{\per}{\mathrm{per}}
\newcommand{\eqr}[1]{~\mbox{$(${\rm \ref{#1}}$)$}}

\newcommand{\Section}[1]{\section{#1}}

\newcommand{\openbox}{\leavevmode
  \hbox to.77778em{%
    \hfil\vrule
  \vbox to.675em{\hrule width.6em\vfil\hrule}%
  \vrule\hfil}} 
\newcommand{\proofname}{Proof}

\newenvironment{proof}[1][\proofname]{\par\normalfont
  \trivlist\item[\hskip\labelsep\itshape #1:]\ignorespaces
  }{\hspace*{1cm}\hspace*{\fill}\openbox \medskip\endtrivlist}

\title{Public Key Cryptography based on Semigroup Actions
\thanks{This work has been supported in part by the Swiss
  National Science Foundation under grant no. 107887. }}%
\date{\today}%
\author{G\'erard Maze \\
  {\small {\em e-mail:\/} gmaze@math.unizh.ch \vspace{-2mm} }\\
  {\small Mathematics Institute\vspace{-2mm}}\\
  {\small University of Zurich\vspace{-2mm}}\\
  {\small Winterthurerstr 190,  CH-8057 Zurich,  Switzerland }\vspace{3mm}
   \and
  Chris Monico \\
  {\small {\em e-mail:\/} cmonico@nd.edu \vspace{-2mm}}\\
  {\small Department of Mathematics and Statistics \vspace{-2mm}}\\
  {\small Texas Tech University \vspace{-2mm}}\\
  {\small Lubbock, TX 79409-1042, USA}\vspace{3mm}
  \and  %
  Joachim Rosenthal \\
  {\small {\em e-mail:\/} rosen@math.unizh.ch \vspace{-2mm} }\\
  {\small Mathematics Institute\vspace{-2mm}}\\
  {\small University of Zurich\vspace{-2mm}}\\
  {\small Winterthurerstr 190,  CH-8057 Zurich,  Switzerland }}

\begin{document}\maketitle
\thispagestyle{empty}
\begin{abstract}
  A generalization of the original Diffie-Hellman key exchange in
  $\left(\Z/p\Z\right)^*$ found a new depth when
  Miller~\cite{mi86} and Koblitz~\cite{ko87a} suggested that such
  a protocol could be used with the group over an elliptic curve.
  In this paper, we propose a further vast generalization where
  abelian semigroups act on finite sets.  We define a
  Diffie-Hellman key exchange in this setting and we illustrate
  how to build interesting semigroup actions using finite
  (simple) semirings. The practicality of the proposed extensions
  rely on the orbit sizes of the semigroup actions and at this
  point it is an open question how to compute the sizes of these
  orbits in general and also if there exists a square root attack
  in general.

  In Section~\ref{Sect:Two} a concrete practical semigroup action
  built from simple semi\-rings is presented. It will require
  further research to analyse this system.
\end{abstract}\vspace{3mm}

\vspace{2mm}
\newpage
\Section{Introduction}                  \label{Sect:Int}

The (generalized) discrete logarithm problem is the basic
ingredient of many cryptographic protocols. It asks the following
question:
\begin{prob}              \label{prob1}
  (see e.g.~\cite{me97}). Given a finite group $G$ and elements
  $g,h\in G$, find an integer $n\in \N$ such that $g^n=h$.
\end{prob}
Problem~\ref{prob1} has a solution if and only if $h\in \langle
g\rangle$, the cyclic group generated by $g$. If $h\in \langle
g\rangle$ then there is a unique integer $n$ satisfying $1\leq
n\leq {\mathrm ord}(g)$ such that $g^n=h$. We call this unique
integer the discrete logarithm of $h$ with base $g$ and we denote
it by $\log_g h$.

Protocols where the discrete logarithm problem plays a
significant role are the Diffie-Hellman key
agreement~\cite{di76}, the ElGamal public key
cryptosystem~\cite{el85}, the digital signature algorithm (DSA)
and ElGamal's signature scheme~\cite{me97}.

The Diffie-Hellman protocol~\cite{di76} allows two parties, say
Alice and Bob, to exchange a secret key over some insecure
channel. In order to achieve this goal Alice and Bob agree on a
group $G$ and a common base $g\in G$. Alice chooses a random
integer $a\in\N$ and Bob chooses a random integer $b\in\N$.
Alice transmits to Bob $g^a$ and Bob transmits to Alice $g^b$.
Their common secret key is $k:=g^{ab}.$

It is clear that solving the underlying discrete logarithm
problem is sufficient for breaking the Diffie-Hellman
protocol. For this reason researchers have been searching for
groups where the discrete logarithm problem is considered a
computationally difficult problem.

In the literature many groups have been proposed as candidates
for studying the discrete logarithm problem. Groups which have
been implemented in practice are the multiplicative group
$(\Z/n\Z)^*$ of integers modulo $n$, the multiplicative group
$\F^*=\F\setminus\{0\}$ of nonzero elements inside a finite field
$\F$ and subgroups~\cite{le00,ru04} of these groups. In recent
time there has been intense study of the discrete logarithm
problem in the group over an elliptic
curve~\cite{bl99,ko87a,mi86,me97} or more generally the group
over an abelian variety~\cite{co06,fr99,ko89}.

In this paper, we show how the discrete logarithm problem over a
group can be seen as a special instance of an action by a
semigroup. The interesting thing is that every semigroup action
by an abelian semigroup gives rise to a Diffie-Hellman key
exchange. With an additional assumption it is also possible to
extend the ElGamal protocol.

The idea of using (semi)group actions for the purpose of building
one-way trapdoor functions is not a new one and it appeared in
one way or the other in several papers. E.g.
Yamamura~\cite{ya98} has been considering a group action of
$Sl_2(\Z)$. Blackburn and Galbraith~\cite{bl99a} have been
analyzing the system of~\cite{ya98} and they have shown that it
is insecure. The key exchange protocol in our paper differs
however from~\cite{ya98} and  the
`bit by bit' computation of Blackburn and Galbraith~\cite{bl99a}
does not apply. Other papers where special instances of semigroup
actions appear are~\cite{an99,ko00a1,sh05,sl07}
and we will say more in a moment.

The paper is structured as follows: In the next section we define
$G$-actions on sets, where $G$ is an arbitrary semigroup. Under
the assumption that $G$ is abelian we define a general
Diffie-Hellman protocol. In Section~3 we consider semigroup
actions which can be linearized in the sense that there exists a
computable homomorphism which embeds the semigroup $G$ into
$\mat_n(\F)$, the ring of $n\times n$ matrices. Section~4 and
Section~5 contain the main results of the paper. We show how
semirings can be used to build interesting abelian semigroup
actions. 

A promising practical example which we are describing in
Section~5 consists of a two sided action. The idea of such an
action originates in the 2003 dissertation of Maze~\cite{ma03t}.
Later, Shpilrain and Ushakov~\cite{sh05} have described similar
two-sided actions in the context of Thompson groups. The
semigroups we are studying in Section~5 are built from simple
semirings.  Simple semirings are of importance as they assure
that the induced matrix semiring is simple.  In the special case
when the semiring is the ring of integers modulo $n$
Slavin~\cite{sl07} filed a patent for the described system citing
the work of Maze. Neither~\cite{sh05} nor~\cite{sl07} build
general semigroup actions starting from semirings.  At this point
it is not clear if there exist parameter ranges where the
described twosided action is simultaneously efficient and
practically secure.

\Section{The generalized Diffie-Hellman protocol}   \label{Sect:Dif}

Consider a semigroup $G$, i.e., a set that comes with an
associative multiplication `$\cdot$'. In particular we do not
require that $G$ has either an identity element or that each
element has an inverse. However, without loss of generality, we
will always assume that the semigroup has an identity. We say
that the semigroup is abelian if the multiplication $\cdot$ is
commutative.

Let $S$ be a finite set and $G$ a a semigroup.
A (left) action of $G$ on $S$ is a map 
\begin{eqnarray*}
  \varphi:\hspace{5mm}G \times S& \longrightarrow &S,
\end{eqnarray*}
satisfying $\phi(g\cdot h, s)=\phi(g, \phi(h, s))$.
We will refer to such an action as a $G$-action on the set $S$,
and when the context is clear, we denote $\phi(g,s)$
simply by $gs$.
Right actions are similarly defined.

We present now the protocols one can define based on semigroup
actions:

\begin{pro}                                      \label{prot1}
  {\bf (Extended Diffie-Hellman Key Exchange)}  Let
  $S$ be a finite set, $G$ be an abelian semigroup, and
  $\phi$ a $G-$action on $S$.  
  The Extended Diffie-Hellman key
  exchange in $(G,S,\phi)$ is the following protocol:
  \begin{enumerate}
  \item Alice and Bob publicly agree on an element $s \in S$.
  \item Alice chooses $a \in G$ and computes $a s$. Alice's
    private key is $a$, her public key is $as$.
  \item Bob chooses $b \in G$ and computes $bs$. Bob's private
    key is $b$, his public key is $bs$.
  \item Their common secret key is then
    $$
    a(bs)=(a\cdot b)s=(b \cdot a)s=b(as).
    $$
  \end{enumerate}
\end{pro}

As in the situation of the discrete logarithm problem it is
possible to construct ElGamal one-way trapdoor functions which
are based on group actions. The interested reader finds more
details in~\cite{ma02p2,mo02t}.

\noindent One would build a cryptosystem based on a semigroup action only if the
following problem is hard:
\begin{prob}  \label{actionprob}
  {\bf (Semigroup Action Problem (SAP)):} Given a semigroup $G$ acting on
  a set $S$ and elements $x\in S$ and $y \in Gx$,
  find $g\in G$ such that $gx=y$.
\end{prob}
If an attacker, Eve, can find an $\alpha \in G$ such that $\alpha
s = as$, then Eve may find the shared secret by computing $\alpha
(bs) = (\alpha \cdot b)s = b(\alpha s) = b(as)$.

Although the semigroup $G$ need not be finite, the finiteness of
$S$ is sufficient in order to provide a bound for the size of the
data during the communication. Nevertheless, if the action
preserves the ``size'' of $s$ with respect to some fixed
representation, finiteness of $S$ is not necessary.

\begin{rem}
  The traditional Diffie-Hellman key exchange 
  is a special instance of Protocol~\ref{prot1}. For this let:
\begin{itemize}
\item $G$ be the semigroup $(\Z, \cdot)$ of integers.
\item $S$ be a cyclic group $H$ where the discrete logarithm
  problem is believed to be difficult.
\item $s$ is a generator of the group $H$ and the action is
  defined by
\begin{eqnarray*}
  \varphi:\hspace{5mm}\Z \times H& \longrightarrow &H\\
(n,s)& \longmapsto &s^n.
\end{eqnarray*}
\end{itemize}
The identity $s^{ab}={(s^a)}^b$ simply says that $\varphi$ is a
commutative $G$-action and the reader readily verifies that
Protocol~\ref{prot1} reduces to the traditional
protocol in this case.
\end{rem}
Of course, there is an analogue version of the Diffie-Hellman
Problem stated in terms of semigroup.

\begin{prob}{\bf (The Diffie-Hellman Semigroup Problem)}   \label{prot2}
  Given a finite abelian semigroup $G$ acting on a finite set $S$
  and elements $x,y,z \in S$
  with $y=g \cdot x$ and $z = h \cdot x$ for some $g,h \in G$,
  find $(gh) \cdot x \in S$.
\end{prob}

The security of Protocol~\ref{prot1} is equivalent to this
problem. The only way we know how to attack Problem~\ref{prot2}
is to solve SAP. It is unknown if SAP and Problem~\ref{prot2} are
equivalent.

\subsection{Generic attacks on the SAP}

First, we should examine the brute force attack.  Suppose Eve
intercepts $as$ and $bs$ through an insecure channel and wants to
decode the ciphertext $a(bs)=b(as)$. She may want to try the
brute force attack to solve Problem~\ref{actionprob}: she
computes $gs$ for all possible $g \in G$ until she finds some
$\alpha$ with $\alpha s = as$. She is then able to break the
system as explained above.  To avoid this attack, Bob and Alice
must choose $G$ and $S$ sufficiently large and select a good
candidate for $s$. Namely, if
$$
G_{Eve} = \{ \alpha \in G \, | \, \alpha s=as \}
$$
then the different parameters $G$, $S$, $s$ must be chosen
such that the size of $G_{Eve}$ is small with respect to the size
of $G$.

If $G$ has the structure of a group (and not just a semigroup)
then $G_{Eve}$ is simply a left coset of the stabilizer group
$$
\mbox{Stab}(s) = \{ g \in G \, | \, g s=s \}
$$
and in this case we are requiring that the quotient group
$G/\mbox{Stab}(s)$ is large.

For a general abelian semigroup $G$ we observe that
$\mbox{Stab}(s)$ is still a sub-semigroup of $G$ and every
element $\alpha\in a\, \mbox{Stab}(s)$ has the property that
$\alpha\in G_{Eve} $, i.e., $ a\, \mbox{Stab}(s)\subset G_{Eve}$.
Again in this case we require that $\mbox{Stab}(s)$ is small in
comparison to $G$.

Note also that every sub-semigroup $H$ of $G$ gives rise to an
equivalence relation on $S$. If one has the ability to
efficiently compute canonical representatives for the equivalence
classes (among other things), this could potentially be used to
an attacker's advantage. But as we will see in
Section~\ref{Sect:Semi}, this is not always an easy task.

It is of course an interesting question if a square root attack
exists for general semigroup actions. In the following we explain
that for special cases this is possible. In general we do
not know how to adapt the known algorithms like e.g. baby step
giant step, or the algorithms Pollard rho or Pollard Kangaroo.

Consider an arbitrary instance of the SAP, where one is given a
semigroup $G$ (say as a subset of $\{0,1\}^N$, with $N$ not too
much larger than $\log_2|G|$), a set $X$ (say as a subset of
$\{0,1\}^M$, with $M$ not too much larger than $\log_2|X|$) and
`black-box' type functions $\pi$ and $\alpha$ for quickly
computing the semigroup product and the action, respectively:
\[
\pi : G\times G\longrightarrow G, \hspace{12pt} \alpha: G\times
X\longrightarrow X.
\]
In addition, one is given $x\in X$ and an element $y\in Gx$ in
the orbit of $x$.  It is also reasonable to assume the
availability of oracles for producing elements of $G$ and $X$
uniformly at random.  The goal then is to find a $g\in G$ for
which $\alpha(g,x)= gx = y$.  We do not know a method for solving
such an arbitrary instance with $O(\sqrt{|G|})$ operations,
except in some special cases.\smallskip

\noindent
\textbf{Situation I:} Suppose an element $g\in G$ is known for
which $g^kx = y$ for some $k\ge 1$. In this case, one first
determines the period and preperiod of $g$ by a method similar to
Pollard's rho method, which needs
$O(\sqrt{\ord(p)})=O(\sqrt{|G|})$ operations, where $\ord(p)$ is
the period plus the preperiod of $g$ (see the definition in
  Section 5). 
  Then the baby-step giant-step method can be applied in an
  obvious way to find $k$ with another
  $O(\sqrt{\ord(p)})=O(\sqrt{|G|})$ operations.  Note: this
  applies immediately to the case where $G$ is a cyclic
  group.\smallskip 

\noindent
\textbf{Situation II:} $G$ is a group, but not cyclic.  For
typical groups, inverses are easily computable, but in any case,
one may always find inverses with $O(\sqrt{|G|})$ group
operations, so it suffices to solve $g_1x=g_2y$, from which one
obtains $(g_2^{-1}g_1)x=y$. For this, a randomized baby-step
giant-step is possible. Compute and store a set $A=\{h_1x,\ldots,
h_mx\}$ for randomly chosen $h_i\in G$ and $m\approx \sqrt{|G|}$.
With clever hashing techniques (or, in the worst case, sorting
$A$) it is possible to quickly test if a given element of $X$ is
in the set $A$.  One then chooses random values of $h\in G$ until
one is found with $hy\in A$. If $hy\in A$, we then have $hy=h_ix$
for some $i$, and so $g=h^{-1}h_i$.\smallskip

If the semigroup is neither a group nor the set-theoretic union
of a small number of cyclic sub-semigroups we do not know how to
adapt the algorithms known for the DLP of abelian groups (see
e.g.~\cite{bu97a}). In contrast to the DLP problem actions of a
semigroup $G$ on a set $X$ can result in a $G$-orbit $Gs$, $s\in
X$, consisting of many ultimately periodic orbits $\{g^ks\mid
k\in\N\}$, $g\in G$. We have observed such phenomena in the
action described in Section~\ref{Sect:Two}. It is an open
research question to come up with a possible square root attack
or to show that under certain conditions a square root attack
cannot exist for general semigroup actions on sets.

For semigroup actions where a square root attack exists and no
other attack is known (like e.g. the DLP over an elliptic curve)
it is generally accepted that an orbit size having 160 bits is
sufficient for practical security. For cases where no square root
attack is known orbit sizes of 80 bits could be sufficient for
practical security.

\Section{Linear abelian semigroup actions over fields}

This section is about linearity in the sense that there is a way
to see the semigroup action as a matrix action on some vector
space. We show that if the correspondence between the two
approaches is computationally feasible, then the Diffie-Hellman
semigroup problem and the semigroup action problem may be solved
easily. Two examples of such action are presented at the end of
the section.

Let us describe the situation more specifically. Let $\F=\F_q$ be the
field with $q$ elements. Suppose we are given an action $G\times S
\longrightarrow S$, with $G$ a finite abelian semigroup and $S$ a
finite set, a semigroup homomorphism $\rho: G \longrightarrow
\mat_n(\F)$ (with multiplication as operation) and an embedding $\psi:
S\longrightarrow \F^n$ such that for all $g\in G,s\in S$ one has
$$
\psi(g \cdot s)=\rho(g)\psi(s).
$$
So $\rho(G)$ is a commutative sub-semigroup of $\mat_n(\F)$.
Let $\F[G]$ be the commutative subalgebra of $\mat_n(\F)$
generated by the elements of $\rho(G)$.

Suppose there exist polynomial time algorithms that compute the
semigroup operation, the semigroup action, the values of the maps
$\rho$ and $\psi$ and polynomial time algorithms that compute
$\rho^{-1}(M)$ for each $M \in \rho(G)$ and $\psi^{-1}(v)$ for each $v
\in \psi(S)$. The next theorem does not take in consideration the
speed of these algorithms. It only describes what can be done at the
level of the linear algebra without taking consideration of the
reduction itself. 
We also suppose we have access to an oracle $\Lambda$ that
allow us to randomly chose elements in $\F[G]$. This assumption takes
into account the desire to capture the situations were the semigroup
$G$ is close to a real matrix algebra.

\begin{thm}\label{lin}
Let $G$, $S$, $\psi$ be arbitrary parameters as above and let
$k=\dim_{\F}\F[G]$. Then:
\begin{enumerate}
\item There exists a probabilistic polynomial time reduction of the
  Diffie-Hellman semigroup problem to a linear algebra problem over
  $\F$ that can be solved in an expected $O(k^2n+n^3)$ number of field
  operations.
\item Let $N= |\F[G]|/|G|$. There exists a probabilistic polynomial
  time reduction of the SAP to a linear algebra problem over $\F$ that
  can be solved in an expected $O(N(k^2n+n^3))$ number of field
  operations.
\end{enumerate}
The above $O$-constants come from the cost of standard linear algebra
problems and bounded expected values.
\end{thm}

\begin{proof}
Let $x$, $y=g \cdot x$ and $z=h \cdot x$ be three elements of $S$ with
$u$,$v$ and $w$ their images in $\F^n$. We consider the semigroup
action problem instance with parameters $x$ and $y$ and the
Diffie-Hellman semigroup problem instance with additional parameter
$z$.
\begin{enumerate}
\item Suppose we have chosen randomly $k$ different elements $M_1,...,$ $M_k$
in $\F[G] \subset \mat_n(\F)$ with $k$ call to the oracle $\Lambda$. 
The probability that this family is in fact a basis of the
vector space $\F[G]$ over $\F$ is equal to the probability $\mathbb{P}$
that a random matrix chosen in $\mat_k(\F)$ is invertible, which
satisfies
\begin{eqnarray}\label{probg}
\mathbb{P} & = &\mbox{Prob}\left(M_1,...,M_k \mbox{ is a basis of }
\F[G] \right)\nonumber \\
& = & \frac{|\mbox{GL}_k(\F)|}{|\mat_k(\F)|} \nonumber \\
& = & \frac{(q^k-1)(q^k-q)...(q^k-q^{k-1})}{q^{k^2}}\nonumber \\
& = & \left(1-\frac{1}{q}\right)\left(1-\frac{1}{q^2}\right)
...\left(1-\frac{1}{q^{k}}\right)\nonumber \\
& > & \prod_{n\geqslant 1} \left(1-\frac{1}{2^n}\right)> 0.28>1/4.
\end{eqnarray}
See \cite{li86} for the cardinality of $\mbox{GL}_k(\F)$. Suppose for the
moment that $\mathcal{B}=\{M_1,...,M_k\}$ is a basis of $\F[G]$. If $k
\geqslant n$ we extract a sub-family of cardinality $n$ say
$M_{i_1},...,M_{i_n}$ of $M_1,...,M_k$ such that
$$
\mbox{Span}_{\F^n}\{M_{i_1}u,...,M_{i_n}u\} =
\mbox{Span}_{\F^n}\{M_{1}u,...,M_{k}u\}.
$$
Note that this is always possible and can be done in $O(k^2n)$ field
operations (see \cite{co93c}). If $k<n$ then we may simply complete
$\mathcal{B}$ with enough zero matrices to have a family of
cardinality $n$. Let us consider the following equations with unknown
$a_1,...,a_n \in \F$ and $b_1,...,b_n \in \F$:
\begin{eqnarray}\label{eqlin}
\left(a_1 M_{i_1} + ... + a_n M_{i_n} \right)u & = & v \nonumber\\
 \mbox{ and }
\;\;\left(b_1 M_{i_1} + ... + b_n M_{i_n} \right)u & = & w.
\end{eqnarray}
If $\mathcal{B}$ is a basis, then both possess at least one solution
because of the property of the family $M_{i_1},...M_{1_n}$. If
$a=[a_1,...,a_n]^t$ and $b=[b_1,...,b_n]^t$ then Equations\eqr{eqlin}
are equivalent to the following :
\begin{eqnarray}\label{eqlin2}
\left[M_{i_1}u \, | \, ... \, | \, M_{i_n}u\right] a & = & v \nonumber\\
\mbox{ and } \;\; \left[M_{i_1}u \, | \, ... \, | \, M_{i_n}u\right] b
& = & w,\nonumber
\end{eqnarray}
and therefore both possess a solution that can be found by solving an $n
\times n$ system of linear equations in $\F$. If the previous systems
do not each have a solution, then we choose another family $M_1,...,M_k$
and restart the process; the number of trials is expected to be less
than 4 by Inequality \ref{probg}. Therefore we can find the vectors $a$
and $b$ in $O(n^3)$ field operations.

The matrices
\begin{eqnarray*}
M_g & = & \left(a_1 M_{i_1} + ... + a_n M_{i_n} \right)\\
 \mbox{ and } \;\;\; M_h & = & \left(b_1 M_{i_1} + ... + b_n M_{i_n} \right)
\end{eqnarray*}
satisfy
$$
M_gM_h = M_hM_g \;,\;\;\;\; M_g u = v \;\;\; \mbox{ and } \;\;\; M_hu =
w.
$$
Let $\sigma = M_gM_hu = M_hM_gu$. Since $M_gu=\rho(g)u$ and
$M_hu=\rho(h)u$, we have
$$
\sigma = M_gM_hu=\rho(g)\rho(h)u= \psi((gh)\cdot x) \Longrightarrow
\psi^{-1}(\sigma) =(gh)\cdot x
$$ which shows that the Diffie-Hellman semigroup problem instance can
be solved after a resolution of a family of problems that take
$O(k^2n+n^3)$ operations over $\F$.
\item The matrix $M_g$ above belongs to $\rho(G)$ with probability $1/N$.
Therefore the number of trials before reaching this state is $O(N)$.
If $M_g \in \rho(G)$, then $\tilde{g} = \rho^{-1} (M_g)$ is a solution
to the semigroup action problem since $\psi(y) = M_g \psi(x) =
\psi(\tilde{g} \cdot x)$.
\end{enumerate}
\end{proof}

Here are some examples where the previous theorem holds or can be used:
\begin{exmp}\label{example41}
Let $M$ be an $n \times n$ matrix with entries in $\F=\F_q$ and
$G=\F[M]$ acting on $\F^n$. If the minimal polynomial of $M$ is $m(x)$
then $\F[M] \cong \F[x]/(m(x))$ (with this isomorphism being efficiently
computable) 
and the latter is a vector space of dimension $k=\mbox{deg }m \leqslant
n$. In such a situation, both the semigroup action problem and
Diffie-Hellman semigroup problem are trivial.
\end{exmp}

\begin{exmp}
  This example comes from invariant theory (see
  e.g.~\cite{st93b2} for an introduction to this classical
  subject). We will consider a contragradient matrix action on
  the ring of polynomials. Fix a finite field $\F = \F_q$, an
  integer $d$ and an abelian sub-semigroup $G$ of $\mat_n(\F)$.
  Let $V_d$ be the vector space over $\F$ of polynomials in
  $\F[x_1,...,x_n]$ of total degree less or equal to $d$. The
  action we are considering is
$$
\begin{array}{ccl}
 G \times V_d & \longrightarrow & V_d \\
 (A,f(x)) & \longmapsto & A \cdot f = f((Ax)^t)\\
\end{array}
$$
where $x=[x_1,...,x_n]^t$ and $Ax$ is the usual matrix multiplication.
This action is linear since $A\cdot (f + g) = A\cdot f + A \cdot g$. If
$r = \dim_{\F} V_d$ then we can naturally embed $V_d$ in $\F^r$ after
having chosen the basis $\mathcal{B}=\{x_1^{e_1}...x_n^{e_n} \; | \;
\sum e_i \leqslant d\}$ of $V_d$. This makes the map $\psi$ easy to
compute and to invert. For sake of clarity, we suppose that
$\mathcal{B}=\{v_1 = x_1,...,v_n=x_n,v_{n+1},...,v_r\}$. We define the
map $\rho : G \longrightarrow \mat_r(\F)$ as follows:
$$
\rho(A)_{ij} = (A \cdot v_j)_i = \left(\prod_{k=1}^{r}
\left(\sum_{l=1}^{n}a_{kl}x_l \right)^{e_k}\right)_i
$$
where $v_j =x_1^{e_1}...x_n^{e_n}$. So $\rho$ gives the matrix
representation of the linear map induced by the action since the
$j^{th}$ column of $\rho (A)$ is the image of the $j^{th}$ basis vector
$v_j$. Since all the polynomials have degree less or equal to $d$, the
right-hand-side can be computed in $O(r n d \log d)$ field operations
(see \cite[Chapter 1]{sh92}). Note that if $M \in \rho(G)$, then we can
easily find $A$ such that $\rho (A)=M$ since the $i^{th}$ row of $A$ is
contained in the $n$ first components of the $i^{th}$ column of $M$.
Indeed, if $1\leqslant i \leqslant n$ then
$$
i^{th} \mbox{ column of M} =A \cdot v_i = \sum_{j=1}^{n}
a_{ij}x_j=\sum_{j=1}^{n} a_{ij}v_j.
$$
Once again the previous theorem holds and makes the
Diffie-Hellman semigroup problem as hard as the linear algebra
problem in $\F^r$. However note that in that case the semigroup
action problem may still be difficult since the ratio $|G|/
|\F[G]|$ may take very small values because of the big dimension
expansion from $n$ to $r$.
\end{exmp}

\Section{Linear actions of abelian semirings on semi-modules}
\label{Sect:Semi}

In this section we construct semigroup actions on finite sets
starting from a semimodule defined over a semiring. 
The setup is general enough that it includes the Diffie-Hellman
protocol over a general finite group as a special case. It
provides on the other hand the flexibility to construct new
protocols where some of the known attacks against the discrete
logarithm problem in a finite group do not work anymore.

Let $R$ be a semiring, not necessarily finite. This means that
$R$ is a semigroup with respect to both addition and
multiplication and the distributive laws hold. It is understood
that the semiring is commutative with respect to addition. Some
authors assume that a semiring has a neutral element with respect
to addition. We will not assume that $R$ has either a zero or a
one.

Let $M$ be a finite semimodule over $R$. With this we mean that
$M$ has the structure of a finite semigroup and there is an
action:
\begin{eqnarray*}
  R\times M &\longrightarrow & M
\end{eqnarray*}
such that
$$
r(sm)=(rs)m,\ (r+s)m=rm+sm \mbox{ and } r(m+n)=rm+rn
$$
for all $r,s\in R$ and $m,n\in M$.

The semigroup action problem in this setting then asks:
\begin{quote}
``Given elements $m,n\in M$ find an element $r\in R$ such that
$rm=n$.
\end{quote}

Before we proceed we would like to explain some of the
difficulties in order to derive at a square root algorithm which
solves the SAP. For this note that many square root attacks seek
in this situation a ``collision'', e.g. in Pollard's rho method
elements $r_1,\ldots,r_4\in R$ are sought such that
\begin{equation}                         \label{coll}
r_1m+r_2n=r_3m+r_4n.
\end{equation}

If the semiring is a ring then this results in
$$ 
(r_1-r_3)m=(r_4-r_2)n
$$
and maybe under benign conditions the semigroup action problem can be
solved. If the semiring (like e.g. the ones we describe in the
next section) have in general no additive inverses this simple
reduction from Equation\eqr{coll} is not possible. The situation
is even worse when $R$ has only a semigroup structure and $M$ is
an arbitrary set since in such a situation no addition is at
disposal at all.

We proceed now and show how to derive at an abelian semigroup action
starting from a semimodule whose coefficient ring is not
necessarily multiplicatively commutative.

Let $\mat_n(R)$ be the set of all $n\times n$ matrices with
entries in the semiring $R$. The semiring structure on $R$
induces a semiring structure on $\mat_n(R)$. Moreover the
semimodule structure on $M$ lifts to a semimodule structure on
$M^n$ via the matrix multiplication:
\begin{eqnarray}      \label{multi}
  \mat_n(R)\times M^n &\longrightarrow & M^n\\
      (A,x) &\longmapsto & Ax.\nonumber
\end{eqnarray}

The action\eqr{multi} forms a semigroup-action of the multiplicative
semigroup of $\mat_n(R)$ on the set $M^n$. In general $\mat_n(R)$ is not
commutative with respect to matrix multiplication. However we can
easily define a commutative subgroup as follows:

Let $C\subset R$ be the center of $R$ i.e., the subset of $R$ consisting of elements
that commute with any other elements.
Let $C[t]$ be the polynomial ring in the indeterminant $t$ and
let $A\in\mat_n(R)$ be a fixed matrix.  If 
$$
p(t)=r_0+r_1t+\cdots + r_kt^k\in C[t]
$$
then we define in the usual way $p(A)=r_0I_n+r_1A+\cdots+r_kA^k$,
where $r_0I_n$ is the $n\times n$ diagonal matrix with entry
$r_0$ in each diagonal element.

Consider the semigroup
$$
G:=C[A]:= \{p(A) \; | \; p(t) \in C[t] \}.
$$
Clearly $C[A]$ has the structure of an abelian semigroup.
Protocol~\ref{prot1} then simply requires that Alice and Bob
agree on a vector $s\in M^n$. Then Alice chooses a matrix $X\in\ 
C[A]$ and sends to Bob the vector $Xs$, an element of the module
$M^n$. Bob chooses a matrix $Y\in C[A]$ and sends to Alice the
vector $Ys$. The common key is then the vector $XYs$ which both
can compute since $X$ and $Y$ commute.

In the special case when $R=M=\F$ is a finite field one readily
reduces the problem to a simple linear algebra problem over the
finite field $\F$. 

The situation becomes slightly more interesting if we take as a
ring $R=\Z$, the integers and as module any finite abelian group
$M=H$.  The group $H$ is a $\Z$ module and $ \mat_n(\Z)$ operates
on $S:=H^n=H\times\ldots\times H$ via the formal multiplication:
\begin{equation}
  \left[ 
  \begin{array}{c}
   g_1 \\ \vdots \\ g_n
  \end{array}
\right]\longmapsto
 \left[ 
  \begin{array}{ccc}
   a_{11}&\ldots&a_{1n}\\ 
   \vdots && \vdots\\  
   a_{n1}&\ldots&a_{nn}
  \end{array}
\right]
  \left[ 
  \begin{array}{c}
   g_1 \\ \vdots \\ g_n
  \end{array}
\right].
\end{equation}
If $l=\lcm\{\vert g_1\vert,\ldots, \vert g_n\vert \}$,
and $C\in \mat_n(\Z)$ is a matrix with all entries congruent to zero
modulo $l$, then $(A+C)g = Ag$ for all $A\in \mat_n(\Z)$. Whence,
we may simply consider the action of $\mat_n(\Z/l\Z)$ on $S$.

This problem reduces to a combination of a linear algebra problem
and a series of discrete logarithm problems in $H$ as soon as all
the elements $\{ g_1, \ldots, g_n\}\subset H$ lie in a common
cyclic subgroup of $H$. Such an attack is even possible when the
$\Z$-action on the abelian group is more complicated and we refer
to the recent system introduced by  Climent et. al.~\cite{cl06}
and its cryptanalysis~\cite{cl07}.

The situation becomes quite a bit more interesting if we consider
general finite semirings acting on general semi-modules. In the
next section we explain an instance where we do not know how to
efficiently attack such a system.

\Section{A two-sided abelian action based on simple semirings}
\label{Sect:Two}

In this section we describe a particular semigroup action, where
we do not know how to solve the SAP once the parameters have been
chosen large enough.  The idea of such an action
originates in the dissertation of Maze~\cite{ma03t}. Shpilrain
and Ushakov~\cite{sh05} have described a similar two-sided action
in the context of Thompson groups and Slavin~\cite{sl07} filed a
patent based on such ideas. 

Let us fix a finite semiring $R$, not embeddable in a field and
not necessarily commutative. Given such a semiring, consider $C$,
the center of $R$. Throughout this section, we let $n$ denote
an arbitrary positive integer.
For $M\in\mat_n(R)$
we denote by $C[M]$ the abelian sub-semiring generated
by $M$, i.e., the semiring of polynomials in $M$ with
coefficients in $C$. Let $M_1,M_2 \in \mat_n(R)$ and consider the
following action:
$$
\begin{array}{ccl}
(C[M_1] \times C[M_2]) \times \mat_n(R) & \longrightarrow & \mat_n(R) \\
((p(M_1),q(M_2)),X)& \longmapsto & p(M_1)\cdot X \cdot q(M_2). \end{array} 
$$
This action is linear since
$$
p(M_1)\cdot (A+B) \cdot q(M_2) = p(M_1)\cdot A \cdot
q(M_2)+p(M_1)\cdot B \cdot q(M_2).
$$
Because of this linearity, we avoid the case when $R$ is a
finite field (see Theorem~\ref{lin}) even if the initial SAP
instance related to this semigroup action looks difficult.

The key-exchange algorithm that results from using this semigroup
action in Protocol \ref{prot1} explicitly reads as follows.
\begin{pro} \label{pro:twosided}
  {\bf (Diffie-Hellman with two-sided matrix semiring action)}
  \label{protDHS}
  \begin{enumerate}
  \item Alice and Bob agree on a finite semiring $R$ with
    nonempty center $C$, not embeddable into a field. They choose
    a positive integer $n$ and matrices $M_1, M_2, S\in
    \mat_n(R)$.
  \item Alice chooses polynomials $p_a,q_a\in C[t]$ and computes
    $A = p_a(M_1)\cdot S\cdot q_a(M_2)$.  She sends $A$ to Bob.
  \item Bob chooses polynomials $p_b, q_b\in C[t]$ and computes
    $B = p_b(M_1)\cdot S\cdot q_b(M_2)$.  He sends $B$ to Alice.
  \item Their common secret key is then
      \[
      p_a(M_1)B q_a(M_2) = p_a(M_1)p_b(M_1) S q_b(M_2)q_a(M_2) =
      p_b(M_1) A q_b(M_2).
       \]
  \end{enumerate}
\end{pro}

The corresponding SAP that should be hard is: given $M_1, M_2,
S\in\mat_n(R)$ and $T\in C[M_1]SC[M_2]$ find $U_1\in
C[M_1]$ and $U_2\in C[M_2]$ so that $T= U_1SU_2$. We do not know
if it is necessary for an attacker to solve this problem,
but it certainly is sufficient.

The remainder of this section is devoted to describing some
necessary conditions on $R$ for this problem to be difficult, and
the existence of semirings meeting these necessary conditions.

\begin{de}
  A congruence relation on a semiring $R$ is an equivalence
  relation $\sim$ such that $a \sim b$ implies that $ ac \sim bc $, 
$ca \sim cb $, $  a +c \sim b+c$ and
  $ c+a \sim c +b $ 
  for all possible choice of $a$, $b$ and $c$. A semiring $R$ is
  congruence-free, or \textit{simple}, if the only congruence
  relations are $R \times R$ and $\{(a,a) \; | \; a \in R \}$.
\end{de}

Any congruence relation induces a natural semiring structure on
the set $R /\hspace{-1.5mm} \sim$ and the quotient map $R
\longrightarrow R /\hspace{-1.5mm} \sim$ is a semiring
homomorphism. It is also clear that a congruence relation on $R$
induces a congruence relation on $\mat_n(R)$ for any $n \in \N$.

For cryptographic purposes it is important that the involved semirings
are simple to avoid a Pohlig-Hellman type reduction of the SAP.
Indeed any congruence relation on $R$ yields a projection of the SAP
instance onto a quotient semiring, from which one may gain information
about the solution to the original instance.  Just as we prefer to
work in groups of prime orders to avoid a Pohlig-Hellman attack, we
would like to work in simple semirings to avoid such a reduction.  Let
us mention that Monico~\cite{mo04} provided a partial classification
of finite simple semirings in 2002 and that Zumbr\"agel recently
provided in \cite{zu07p} a total classification of non-trivial finite
simple semirings together with a method for explicitly constructing
such objects. For this we first define:
\begin{de}
  A zero of a semiring $R$ is an element `0' such that
  $a+0=0+a=a$ and $a\cdot 0 = 0\cdot a=0$ for all $a\in R$. A one
  of a semiring $R$ is an element `1' such that $a\cdot 1 =
  1\cdot a=a$ for all $a\in R$.
\end{de}

Next we show how to build large simple semirings from small
simple semirings. We start with a technical lemma:

\begin{lem}
  Let $R$ be an additively commutative semiring with $1$ and $0$
  and let $\sim$ be a congruence relation on $\mat_n(R)$. Then
  there exists a congruence relation $\sim_0$ on $R$ such that
  $$
  A \sim B \in \mat_n{R} \Longleftrightarrow a_{ij} \sim_0
  b_{ij}\;\;, \; \forall\; 0 \leqslant i,j \leqslant n.
  $$
\end{lem}

\begin{proof}
  First, given such a semiring $R$, and $M \in \mat_n(R)$, if
  $M'$ is obtained from $M$ by a permutation of rows and columns,
  we prove there exist two invertible matrices $S,P \in
  \mat_n(R)$ such that $M'= SMP$. Indeed, the statement is true
  if one consider matrices with entries in $\Z$ and the usual
  multiplication, i.e., there exist two permutation matrices
  (therefore with entries in $\{0,1\}$) such that $M'= S\cdot M
  \cdot P$ with $\cdot$ being the usual matrix multiplication. It
  is then straightforward to verify that the same is true with
  the operation in $R$ because of the properties of $0$ and $1$.
  Let us now prove the stated result.  Let $f:R \longrightarrow
  \mat_n(R)$ be the map that sends $a \in R$ to the diagonal
  matrix with first diagonal element $a$ and zeros everywhere
  else. The map $f$ is a semiring homomorphism. Let $\sim_0$ be
  the relation on $R$ defined by $a \sim_0 b$ in $R$ if and only
  if $f(a) \sim f(b)$ in $\mat_n(R)$. Observe that $\sim_0$ is a
  congruence relation on $R$. We prove now that the statement of
  the lemma is true for $\sim_0$. Let $A,B \in \mat_n(R)$ and
  $J=f(1)$. Let $0 \leqslant i,j \leqslant n$ and $S_{ij},P_{ij}
  \in \mat_n(R)$ be permutation matrices such that
  $$
  (S_{ij}AP_{ij})_{11}=a_{ij} \;\;\mbox{ and } \;\;
  (S_{ij}BP_{ij})_{11}=b_{ij}.
  $$
  Note that the matrices $S_{ij}$ and $P_{ij}$ exists in
  $\mat_n(R)$ by the previous remark.  Therefore $JS_{ij}AP_{ij}J
  =
  f(a_{ij})$ and $JS_{ij}BP_{ij}J = f(b_{ij})$.\\
  `$\Longrightarrow$': If $A \sim B$ then
  $JS_{ij}AP_{ij}J \sim JS_{ij}BP_{ij}J$ and
  therefore $a_{ij} \sim_0 b_{ij}$.\\
  `$\Longleftarrow$': Clearly
  $$
  A = \sum_{i,j} S_{ij}^{-1}f(a_{ij})P_{ij}^{-1} \mbox{ and }
  B = \sum_{i,j} S_{ij}^{-1}f(b_{ij})P_{ij}^{-1}
  $$
  and since $f(a_{ij}) \sim f(b_{ij})$, $A\sim B$.
\end{proof}

As an immediate consequence of this lemma, we have the following
theorem which provides arbitrarily large, finite, simple semirings.
\begin{thm}\label{c-s}
  Let $R$ be an additively commutative semiring with $1$ and $0$
  and let $n \in \N$. Then $R$ is simple if and only if $\mat_n(R)$ is
  simple.
\end{thm}

With the help of this Theorem we can readily build large finite
simple semirings with 0,1 which are not rings and not embeddable
into fields.  The following provides several explicit examples of
some small finite simple semirings with 0,1 which are not rings
and not embeddable into fields.

\begin{exmp}
  Consider the set $S=\{0,1\}$ with the operations $\max$ and
  $\min$ for addition and multiplication respectively.  One
  readily verifies that $S$ has the structure of a finite simple
  semiring.  Note that several polynomial time problems over
  $\Z$, such as polynomial factorization, have been found to be
  NP-hard when considered over this semiring $S$ \cite{ki05u}.
\end{exmp}

The following example was found by computer search.

\begin{exmp}
  Consider the set $S_{6,1}=\{0,1,2,3,4,5\}$ satisfying the
  following addition and multiplication rules.
\begin{center}
\begin{tabular}{c||c|c|c|c|c|c|}
+  &     0   &   1  & 2&         3   &   4   &   5 \\ \hline\hline
0   &   0&      1&2&    3&      4&      5\\ \hline
1   &   1&      1&1&    1&      1&      5\\ \hline
2   &   2&      1&2&    1&      2&      5\\ \hline
3   &   3&      1&1&    3&      3&      5\\ \hline
4   &   4&      1&2&    3&      4&      5\\ \hline
5   &   5&      5&5&    5&      5&      5\\ \hline
\end{tabular} 
\hspace{1cm}
\begin{tabular}{c||c|c|c|c|c|c|}                        
$\cdot$ &       0  &    1   &   2   &   3  &    4  &    5  \\ \hline\hline
  0   & 0&      0&   0& 0 &     0&      0\\ \hline
1   &   0&      1& 2&   3&      4&      5\\ \hline
2   &   0&      2& 2&   0&      0&      5\\ \hline
3   &   0&      3& 4&   3&      4&      3\\ \hline
4   &   0&      4& 4&   0&      0&      3\\ \hline
5   &   0&      5& 2&   5&      2&      5\\ \hline
\end{tabular}
\end{center}
$S_{6,1}$ is a finite simple semiring with 6 elements. This is up
to isomorphism the only simple semiring of order 6. This result
follows from~\cite{mo04,zu07p}. 
\end{exmp}

\begin{exmp}              \label{exempZu}
  Using the classification of J.~Zumbr\"agel derived
  in~\cite{zu07p} it is possible to derive for many orders
  addition and multiplication tables.  We are grateful to
  J.~Zumbr\"agel for providing us with the following recently
  found simple semiring having order 20.  Details on how to
  construct the addition and multiplication table can be found
  in~\cite{zu07p}. Again one can show that this is up to
  isomorphism the only simple semiring of order 20.

\renewcommand{\arraystretch}{0.65} \tabcolsep 0.6mm
\begin{center}
\begin{tabular}{c|cccccccccccccccccccc}
+ &0&a&b&c&d&e&f&g&h&i&j&k&l&m&n&o&p&q&r&1  \\ \hline   
0 &0&a&b&c&d&e&f&g&h&i&j&k&l&m&n&o&p&q&r&1  \\     
a &a&a&b&c&d&e&f&g&h&i&j&k&l&m&n&o&p&q&r&1   \\     
b &b&b&b&c&e&e&f&g&h&i&k&k&l&m&n&o&p&q&r&1   \\     
c &c&c&c&c&f&f&f&h&h&i&l&l&l&1&n&p&p&q&r&1   \\     
d &d&d&e&f&d&e&f&g&h&i&j&k&l&m&n&o&p&q&r&1   \\     
e &e&e&e&f&e&e&f&g&h&i&k&k&l&m&n&o&p&q&r&1   \\     
f &f&f&f&f&f&f&f&h&h&i&l&l&l&1&n&p&p&q&r&1   \\     
g &g&g&g&h&g&g&h&g&h&i&m&m&1&m&n&o&p&q&r&1  \\      
h &h&h&h&h&h&h&h&h&h&i&1&1&1&1&n&p&p&q&r&1   \\     
i &i&i&i&i&i&i&i&i&i&i&n&n&n&n&n&q&q&q&r&n   \\     
j &j&j&k&l&j&k&l&m&1&n&j&k&l&m&n&o&p&q&r&1   \\     
k &k&k&k&l&k&k&l&m&1&n&k&k&l&m&n&o&p&q&r&1   \\     
l &l&l&l&l&l&l&l&1&1&n&l&l&l&1&n&p&p&q&r&1   \\     
m &m&m&m&1&m&m&1&m&1&n&m&m&1&m&n&o&p&q&r&1   \\    
n &n&n&n&n&n&n&n&n&n&n&n&n&n&n&n&q&q&q&r&n   \\    
o &o&o&o&p&o&o&p&o&p&q&o&o&p&o&q&o&p&q&r&p   \\     
p &p&p&p&p&p&p&p&p&p&q&p&p&p&p&q&p&p&q&r&p   \\     
q &q&q&q&q&q&q&q&q&q&q&q&q&q&q&q&q&q&q&r&q   \\     
r &r&r&r&r&r&r&r&r&r&r&r&r&r&r&r&r&r&r&r&r   \\     
1 &1&1&1&1&1&1&1&1&1&n&1&1&1&1&n&p&p&q&r&1   \\
\end{tabular}  \smallskip\tabcolsep 0.74mm

\begin{tabular}{c|cccccccccccccccccccc}
$\cdot$ &0&a&b&c&d&e&f&g&h&i&j&k&l&m&n&o&p&q&r&1 \\ \hline   
0 &0&0&0&0&0&0&0&0&0&0&0&0&0&0&0&0&0&0&0&0 \\
a &0&0&0&0&0&0&0&0&0&0&a&a&a&a&a&b&b&b&c&a \\
b &0&0&0&0&a&a&a&b&b&c&a&a&a&b&c&b&b&c&c&b \\
c &0&a&b&c&a&b&c&b&c&c&a&b&c&b&c&b&c&c&c&c \\
d &0&0&0&0&0&0&0&0&0&0&d&d&d&d&d&g&g&g&i&d \\
e &0&0&0&0&a&a&a&b&b&c&d&d&d&e&f&g&g&h&i&e \\
f &0&a&b&c&a&b&c&b&c&c&d&e&f&e&f&g&h&h&i&f \\
g &0&0&0&0&d&d&d&g&g&i&d&d&d&g&i&g&g&i&i&g \\
h &0&a&b&c&d&e&f&g&h&i&d&e&f&g&i&g&h&i&i&h \\
i &0&d&g&i&d&g&i&g&i&i&d&g&i&g&i&g&i&i&i&i \\
j &0&0&0&0&0&0&0&0&0&0&j&j&j&j&j&o&o&o&r&j \\
k &0&0&0&0&a&a&a&b&b&c&j&j&j&k&l&o&o&p&r&k \\
l &0&a&b&c&a&b&c&b&c&c&j&k&l&k&l&o&p&p&r&l \\
m &0&0&0&0&d&d&d&g&g&i&j&j&j&m&n&o&o&q&r&m \\
n &0&d&g&i&d&g&i&g&i&i&j&m&n&m&n&o&q&q&r&n \\
o &0&0&0&0&j&j&j&o&o&r&j&j&j&o&r&o&o&r&r&o \\
p &0&a&b&c&j&k&l&o&p&r&j&k&l&o&r&o&p&r&r&p \\
q &0&d&g&i&j&m&n&o&q&r&j&m&n&o&r&o&q&r&r&q \\
r &0&j&o&r&j&o&r&o&r&r&j&o&r&o&r&o&r&r&r&r \\
1 &0&a&b&c&d&e&f&g&h&i&j&k&l&m&n&o&p&q&r&1 \\
\end{tabular}  
\end{center}
\end{exmp}
\renewcommand{\arraystretch}{1}

In order that the two-sided semigroup action described in the
beginning of this section is difficult we would like that the
sets $C[M_1]$ and $C[M_2]$ are large with regard to the matrix
size $n$.  The orders of the matrices $M_1$ and $M_2$ chosen to
act on the matrix $A$ on the left and on the right are of prime
importance.  Indeed the cardinality of the commutative semiring
$C[M]$ directly depends on the order of $M$. We study the
``sizes'' of the orbit of powers of elements in $\mat_n(S)$ where
$S=\{\{0,1\}, \max, \min\}$. We will see that these orders give
lower bounds for the maximum orders of elements in any semiring
with 0 and 1.  Note that since the semiring $\mat_n(S)$ is finite any
sequence $\{M^k\}_{k \in \N}$ will eventually repeat, i.e.,
create a collision of the form $M^k=M^{k'}$ with $k\ne
k'$. Computer experiments also showed that in general the set
$C[M]$ is much larger than the set $M^k=M^{k'}$.

\begin{de}\label{deord}
  Let $a=\{a_k\}_{k \in \N}$ be a sequence in a finite set such that
  $a_n=a_m \Longrightarrow a_{n+1} = a_{m+1}$. The \textit{order}
  $\ord(a)$ of $a$ is the least positive integer $m$ for which
  there exists $k< m$ with $a_k=a_m$. The
  \textit{preperiod} $p_r(a)$ of $a$ is the largest non-negative
  integer $m$ such that for all $k>m$ we have $a_k \neq a_{m}$.
  The \textit{period} $\per(a)$ of $a$ is the least positive integer
  $m$ for which there exists an integer $N$ with $a_{m+k}=a_{k}$
  for all $k>N$. If $g$ is an element of a semigroup, then we set
  $\ord(g)=\ord(\{g^n\}_{n \in \N})$, $\per(g)=\per(\{g^n\}_{n \in
    \N})$ and $p_r(g)=p_r(\{g^n\}_{n \in \N})$.
\end{de}

Clearly $\ord(a)=\per(a)+p_r(a)$. Returning to the situation of the
multiplicative semigroup of $\mat_n(S)$, we study the question ``How
large can the order of $M \in \mat_n(S)$ be?''. There already exist
some results in this direction. To describe them, we recall that
for a given oriented graph $G$, a \textit{strongly connected
  component} (written SCC) of $G$ is a sub-graph $H$ of $G$
inside which any two vertices $i$ and $j$ belong to a common
oriented cycle and $H$ is a maximal sub-graph with this property.
Such a SCC is written $H\subseteq_{SCC} G$. The period of a
strongly connected component is the maximum between the $\gcd$ of
the length of its cycles and 1. We refer the reader to
\cite{li95} for the details.

\begin{pr}\label{gav}
  Let $M \in \mat_n(S)$ and $G$ be the directed graph whose adjacency
  matrix is $M$. Then
\begin{enumerate}
\item $\per(M) = \lcm \{ \mbox{period of $H$ }| \; H \mbox{ is a SCC
    of }G\}$,
\item The numbers $\per(M),p_r(M)$ and $\ord(M)$ can be computed in
  $O(n^3)$ time.
\end{enumerate}
\end{pr}

This proposition is essentially in \cite{ga97b}. The algorithm
given there computes $\per(M)$ in $O(n^3)$ time and an easy
modification of it allows to computes $p_r(M)$ and therefore
$\ord(M)$.

We introduce now a function that play a crucial role: Landau's
function $g$. It is defined by
\begin{eqnarray*}
g(n) & = & \max\{ \ord (\sigma)\; | \; \sigma \in S_n\} \\
& = & \max\{\lcm\{a_1,...,a_m\}\; | \; a_i >0,\; a_1+...+a_m=n\}.
\end{eqnarray*}
It was first studied by Landau \cite{la03} in 1903 who proved
that
\begin{equation}\label{landau}
\ln(g(n))\sim \sqrt{n\ln (n)} \;\;\;\mbox{ as } \;\;n \longrightarrow
\infty.
\end{equation}
In 1984, Massias \cite{ma85} showed that for sufficiently large
$n$,
\begin{equation}\label{massias}
\sqrt{n \ln (n)} \leqslant \ln(g(n)) \leqslant \sqrt{n \ln (n)}
\left(1+ \frac{\ln \ln (n)}{2 \ln(n)}\right),
\end{equation}
the second inequality in~\ref{massias} being true for all $n$.
Clearly, the function $g$ is increasing.  In any case, we have
$$
\max\{\lcm\{a_1,..,a_m\} : |a_1|+...+|a_m|=n\} = \exp\left(
  (1+o(1)) \sqrt{n\ln n} \right).
$$
On the other hand, the period of any SCC $H \subset G$ is less
or equal to $|H|$ and
$$
\sum_{\substack{H \subseteq_{SCC} G}} |H| \leqslant n.
$$
Since the function $g$ is increasing, Proposition \ref{gav}
and Equation\eqr{landau} give
$$
\per(M) \leqslant g\left( \sum_{\substack{H \subseteq_{SCC} G}}
  |H| \right) \leqslant g(n) = \exp\left( (1+o(1)) n^{1/2}
  \ln^{1/2} n \right).
$$
Further, it is not difficult to see that there always exists
an oriented graph $G$ with period $g(n)$. Indeed if $g(n)$ is
reached by a partition $a_1+...+a_m=n$, then a graph $G$ built
out of cyclic SCCs of order $a_i$ satisfies $\per(M)=g(n)$.  Such a
matrix $M \in \mat_n(S)$ that reaches this bound is in fact a
permutation matrix, and as such, it can be seen as an element of
any semirings with 0 and 1. In other words, in any such semiring,
the previous bound is reached:

\begin{pr}\label{ordg}
  Let $n \in \N$ and $R$ be a semiring with 0 and 1. Then
  $$
  \max\{\per(M) \; | \; M \in \mat_n(R)\} \geq g(n) =\exp\left(
    (1+o(1)) n^{1/2} \ln^{1/2} n \right).
  $$
  If $R=S=\{\{0,1\},\max,\min\}$, then the above inequality is
  an equality.
\end{pr}

The exact computation of $g(n)$, or more precisely, of the
partition $a_1+...+a_m=n$ that yields the maximum $g(n)$, is
necessary in order to build explicitly a matrix $M \in \mat_n(S)$ such
that $\per(M)=g(n)$.  Indeed, the integer $g(n)$ is always a product
of primes less or equal to $2.86 \sqrt{n \ln (n)}$, c.f.
\cite{ma89b}. Therefore the factorization of $g(n)$ can be found
in polynomial time in $n$. It is also known that the partition of
$n$ that gives the maximum $\lcm$ has parts that are all prime
powers, c.f. \cite{gr95}, and therefore the factorization of
$g(n)$ gives the expected partition directly. The algorithm given
in \cite{ni69} allows one to compute $g(n)$ for large integers
$n$, up to $n=32,000$, so the exact determination of the matrix
$M$ is not a problem. See Table 5.1 for a list of
values of $g(n)$ with the associated partition. 

\begin{table}[h]\caption{Some values of Landau's function $g$}
\begin{center}
\begin{tabular}{|c|l|l|}
  \hline
  $n$ & $g(n)$ & Associated partition \\ \hline
  256 & 4243057729190280 & 8, 9, 5, 7, 11, 13, 17, 19, 23,\\
  &&  29, 31, 41, 43 \\ \hline
  512 & 70373028815644182 $\backslash$ & 1, 1, 1, 4, 9, 5, 7, 11, 13, 17, \\
  &5899620 &  19, 23, 29, 31, 37, 41, 43, 47,\\
  & &  53, 59, 61 \\ \hline
  1024 & 855674708268439827 $\backslash$ & 1, 1, 1, 16, 27, 25, 7, 11, 13,\\
  & 7434193536488991600 &  17, 19, 23, 29, 31, 37, 41, 43, \\
  &&      47, 53, 59, 61, 67, 71, 73, 79,  \\
  &&  83, 89 \\ \hline
\end{tabular}
\end{center}
\end{table}

For a given matrix $M \in \mat_n(S)$, since $S[M] \supset \{M^k\}_{k
  \in \N}$, we have
$$
|S[M]| \geqslant \ord(M) \geqslant \per(M),
$$
and the last inequality can give $|S[M]| \geqslant g(n)$ for a
wisely chosen $M$.

The following corollary shows that the size of the sets $C[M]$
grows exponentially in $n$ for suitable matrices $M$ as soon as
the center $C$ contains the elements $0,1$ of a semiring. Such
matrices can even be constructed in an efficient way.

\begin{co}
  Let $n \in \N$ and $R$ be a semiring with 0 and 1 and center
  $C$. Then there is an $n\times n$ matrix $M$ with entries in
  $R$ such that the order of $M$ is larger than $g(n)$ in
  particular the size of $C[M]$ is larger than $g(n)$ as well.
\end{co}

We conclude the Section with an example to illustrate how finite
simple semirings could be used to build a practical semigroup
action problem.

\begin{exmp}
  Consider the semiring $R=S_{6,1}$ as defined above. The
  elements $\{0,1\}$ form the center $C$ of $R$.  We will
  consider the matrix ring $\mat_n(R)$ with $n=20$. In this
  situation the key size is $400 \cdot \lg 6 \cong 1033$ bits and
  the value of Laudau's $g$ function is $g(20)=1 \cdot 4 \cdot 3
  \cdot 5 \cdot 7 = 420$. By the last corollary $\mat_n(R)$
  contains elements $M$ whose multiplicative order $\ord(M)$ is
  at least $420$.  For such an element $M$ the abelian semigroup
  $C[M]$ contains all elements of the form $\sum_{i=0}^k r_iM^i$
  with $r_i\in\{0,1\}$. The size of the set $C[M]$ is upper
  bounded by $2^{k+1}$, where $k=\ord(M)$.

 The matrices $M_1$ and $M_2$ below  are chosen to be close to
 permutation matrices such that the orders are actually more than
 420. The matrix $S$ is also chosen sparse as computer
 experiments with the particular ring $S_{6,1}$ showed that this
 leads to maximal possible size of the possible matrices 
$$
C[M_1] \cdot S \cdot C[M_2].
$$
Upon using these parameters in
  Protocol \ref{pro:twosided}, Alice chooses polynomials
  $p,q\in C[t]$ and computes
  $$
    A:=p(M_1) \cdot S \cdot q(M_2)
  $$
  $p, q\in C[t]$ were chosen as private keys by Alice in
  Protocol~\ref{pro:twosided}.

  It is clear that she has more than $2^{420}$ choices to choose
  a polynomial  $p\in  C[t]$ and for such a polynomial $p(M_1)$
  can be computed with at most 420 matrix multiplication and
  addition. - Of course Alice can restrict herself to polynomials
  of smaller degree, say e.g. $k<50$ which leaves still $2^{50}$
  choices for $p$ and for $q$ and which reduces the number of
  matrix multiplications and additions to 100, a task
  quite easy for an average PC. 

  Assume Alice has chosen the matrices in the following
  particular way:
 
\renewcommand{\arraystretch}{0.5} \tabcolsep 0.3mm
{\small
  $$
  \!\!\! M_1\! =\!\! \left[
\begin{tabular}{cccccccccccccccccccc}
1&0&0&0&0&0&0&0&0&0&0&0&0&0&0&0&0&0&0&0\\
0&0&1&0&0&0&0&0&0&0&0&1&0&0&0&0&0&0&0&0\\
0&0&0&1&0&0&0&0&0&0&0&0&0&0&0&0&0&0&0&0\\
0&0&0&0&1&0&0&0&0&0&0&0&0&0&0&0&0&0&0&0\\
0&1&0&0&0&0&0&0&0&0&0&0&0&0&0&0&0&0&0&0\\
0&0&0&0&0&0&1&0&0&0&0&0&0&0&0&0&0&0&0&0\\
0&0&0&0&0&0&0&2&0&0&0&0&0&0&0&0&0&0&1&0\\
0&0&0&0&0&1&0&0&0&0&0&0&0&0&0&0&0&0&0&0\\
0&0&0&0&0&0&0&0&0&1&0&0&0&0&0&0&0&0&0&0\\
0&0&0&0&0&0&0&0&0&0&1&0&0&0&0&0&0&0&0&0\\
0&0&0&0&0&0&0&0&0&0&0&2&0&0&0&0&0&0&0&0\\
0&0&0&0&0&0&0&0&0&0&0&0&1&0&0&0&0&0&0&0\\
0&0&0&0&0&0&0&0&1&0&0&0&0&0&0&0&0&0&0&0\\
0&0&0&0&0&0&0&0&0&0&0&0&0&0&1&0&0&0&0&0\\
0&0&0&0&0&0&0&0&0&0&0&0&0&0&0&1&0&0&0&0\\
0&0&0&0&0&0&0&0&0&0&0&0&0&0&0&0&1&0&0&0\\
0&0&0&1&0&0&0&0&0&0&0&0&0&0&0&0&0&1&0&0\\
0&0&0&0&0&0&0&0&0&0&0&0&0&0&0&0&0&0&1&0\\
0&0&0&0&0&0&0&0&0&0&0&0&0&0&0&0&0&0&0&1\\
0&0&0&0&0&0&0&0&0&0&0&0&0&1&0&0&0&0&0&0
\end{tabular}
\right] \; M_2\!=\! \left[
\begin{tabular}{cccccccccccccccccccc}
0&0&0&0&0&0&0&0&0&0&0&0&0&0&0&0&0&0&1&0\\
0&0&0&0&0&0&0&0&0&0&0&1&0&0&0&0&0&0&0&0\\
0&0&0&0&0&0&1&0&0&0&0&0&0&0&0&0&0&0&0&0\\
0&0&1&0&0&0&0&0&0&0&1&0&0&0&0&0&0&0&0&0\\
0&0&0&0&0&0&0&0&0&0&0&0&0&0&0&0&0&0&0&4\\
0&0&0&0&0&0&0&0&0&0&0&0&0&0&0&1&0&0&0&0\\
0&2&0&0&0&0&0&0&0&0&0&0&0&0&0&0&0&0&0&0\\
0&0&0&0&0&0&0&0&0&0&0&0&0&0&0&0&0&1&0&0\\
0&0&0&1&0&0&0&0&1&0&0&0&0&0&0&0&0&0&0&0\\
0&0&0&0&0&0&0&0&0&0&0&3&1&0&0&0&0&0&0&0\\
0&0&0&0&0&0&0&0&0&0&0&0&0&0&2&0&0&0&0&0\\
0&0&0&1&0&0&0&0&0&0&0&0&0&0&0&0&0&1&0&0\\
0&0&0&0&0&0&0&0&0&0&1&0&0&0&0&0&0&0&0&0\\
0&0&0&0&0&1&0&0&0&0&0&0&0&0&0&0&0&0&0&0\\
0&0&0&0&0&0&0&0&0&1&0&0&0&0&0&0&0&0&0&0\\
0&0&0&0&0&0&0&1&0&0&0&0&0&0&0&0&0&0&0&0\\
1&0&0&0&0&0&0&0&0&0&0&0&0&0&0&0&0&0&0&0\\
0&0&0&0&1&0&0&0&0&0&0&0&0&0&0&0&0&0&0&0\\
0&0&0&0&0&0&2&0&0&0&0&0&0&0&0&0&1&0&0&0\\
0&0&0&0&0&0&0&0&0&0&0&0&0&1&0&0&0&0&0&0
\end{tabular}
\right]
$$

$$
S= \left[
\begin{tabular}{cccccccccccccccccccc}
0&1&0&0&0&0&0&0&0&0&0&0&0&0&0&0&0&0&0&0\\
0&0&1&0&0&0&0&0&0&0&0&1&0&0&0&0&0&0&0&0\\
1&0&0&0&0&0&0&0&0&0&0&0&0&0&0&0&0&0&0&0\\
0&0&0&1&0&1&0&0&0&0&0&0&0&0&0&0&0&0&0&0\\
0&0&0&0&0&0&0&1&0&0&0&0&0&0&0&0&1&0&0&0\\
0&0&0&0&0&0&1&0&0&0&0&0&0&0&0&0&0&0&0&0\\
0&0&0&0&0&1&0&0&0&0&0&0&0&0&0&0&0&0&0&0\\
0&1&0&0&0&0&0&1&0&0&0&0&0&0&0&1&0&0&0&0\\
0&0&0&0&0&0&0&0&0&1&0&0&0&0&0&1&0&2&0&0\\
0&0&0&0&0&0&0&0&0&0&1&0&0&0&0&0&0&0&0&0\\
0&0&0&0&0&0&0&0&0&0&0&1&0&0&0&0&0&0&0&0\\
0&0&0&0&0&0&0&0&1&0&0&0&0&0&0&0&0&0&0&0\\
0&0&0&1&0&0&0&0&0&0&0&0&1&0&0&0&0&0&0&0\\
0&0&0&0&0&0&0&0&0&0&0&0&5&1&0&0&0&0&0&0\\
0&0&1&0&0&0&0&0&0&0&0&0&0&0&1&0&0&0&0&1\\
0&0&0&1&0&0&0&0&0&0&0&0&0&0&0&1&0&0&0&0\\
0&0&0&0&0&0&0&0&0&2&0&0&0&0&0&0&0&0&0&1\\
0&0&2&0&0&0&0&0&0&0&0&0&1&0&0&0&0&0&1&0\\
0&0&0&0&1&0&0&0&0&0&0&0&0&0&0&0&0&1&0&0\\
0&0&0&1&0&0&0&0&0&0&0&0&0&0&0&0&1&0&0&0
\end{tabular}
\right] \;\;\; A\!=\! \left[
\begin{tabular}{cccccccccccccccccccc}
0&1&2&2&2&0&2&4&0&2&2&2&2&4&2&4&0&2&0&0\\
 1&2&1&1&2&1&1&1&1&1&1&1&1&4&2&1&1&2&1&4\\
 1&2&1&1&2&1&1&1&1&1&1&1&1&4&2&1&1&2&1&4\\
 1&2&1&1&2&1&1&1&1&1&1&1&1&4&2&1&1&2&1&4\\
 1&2&1&1&2&1&1&1&1&1&1&1&1&4&2&1&1&2&1&4\\
 1&2&2&1&1&1&2&1&0&2&2&2&5&1&2&1&1&1&1&1\\
 1&2&1&1&1&1&1&2&0&2&2&1&5&1&1&2&1&1&1&1\\
 1&2&1&1&1&1&1&2&0&2&2&2&1&4&1&1&1&1&1&1\\
 0&2&2&2&2&4&2&4&2&2&1&1&2&0&2&0&0&2&0&0\\
 0&2&2&2&2&4&2&4&2&2&2&1&2&0&2&2&0&2&0&0\\
 0&2&2&2&2&0&2&0&2&2&2&2&2&0&2&2&0&2&0&0\\
 0&2&2&1&2&4&2&4&0&1&1&1&1&4&2&1&0&2&0&0\\
 0&2&2&2&2&4&2&4&2&1&1&1&2&4&2&1&0&2&0&0\\
 1&2&1&1&1&1&1&1&1&1&1&1&5&1&1&1&1&1&1&1\\
 1&2&1&1&1&1&1&1&1&2&1&1&5&1&1&1&1&1&1&1\\
 1&2&1&1&1&1&1&1&1&1&1&1&1&4&1&1&1&1&1&1\\
 1&2&1&1&1&1&1&1&1&1&1&1&5&1&2&1&1&1&1&1\\
 1&2&1&1&1&1&1&4&0&2&2&1&5&1&1&4&1&1&1&1\\
 1&2&1&1&1&1&1&1&1&2&1&2&5&1&1&1&1&1&1&1\\
 1&2&1&1&1&1&1&1&0&2&2&1&5&1&1&1&1&1&1&1
\end{tabular}
\right]
$$}
\renewcommand{\arraystretch}{1}
  
  The only way we know for an attacker to break this system would
  be to find polynomials $\tilde{p}$ and $\tilde{q}$ such that
  $\tilde{p}(M_1)S\tilde{q}(M_2)=A$ (or, to solve a similar
  problem in terms of the matrix $B$ Bob computes). If the
  degrees of $p,q$ are in the range of $50$ a brute force search
  will depend on the size of the set:
$$
\mathcal{S}:=\{p(M_1) \cdot S \cdot q(M_2)\mid \deg
  p<50, \deg q<50\}.
  $$
  An immediate upper bound for the size of the set
  $\mathcal{S}$ is $2^{100}$.  We did run extensive computations
  and could show that $\mathcal{S}$ has size at least $2^{25}$,
  not sufficient to be used as a practical system.  It will
  require further research to estimate better the size of
  $\mathcal{S}$ and to understand how the sizes grow as we
  increase both the matrices involved and the simple semirings.
  E.g. one could run the protocol with the semiring of
  Example~\ref{exempZu} and leave the size of the matrices the
  same. 

  In order to describe the efficiency of the system assume that 
  Alice and Bob agree on matrices of size $n$, polynomials $p,q$ of
  degree at most $k$ and a simple semiring $R$ of cardinality
  $|R|=\theta$. Then the public key and the data to be
  transmitted has $O(n^2\lg \theta)$ bits.
  The number of required bit operations during encryption is
  $O(kn^3(\lg\theta))$ and the computation of the common secret
  key requires  $O(n^3(\lg\theta))$ bit operations. If
  $\tilde{\theta}$ denotes the cardinality of the center $C$ of
  $R$ then an upper bound for the size of the set $\mathcal{S}$
  is $\tilde{\theta}^{2k}$. 

  These complexity estimates suggest that the system should be
  further analysed in particular when the sizes of the matrices
  are small and the sizes of the ring $R$ is large.
\end{exmp}

\Section{Conclusion}

An abelian group can be viewed in a natural way as $\Z$-module.
In this paper we consider the situation when an arbitrary
semigroup (instead of just the integers) act on an arbitrary
finite set. The generalization of the discrete logarithm problem
results in the semigroup action problem which we study in this
paper.  In the situation when the semigroup is abelian one has a
natural Diffie-Hellman secret key exchange and a sufficient
condition to break the key exchange is to solve the semigroup action
problem.

In the later part of the paper we concentrate on a particular
semigroup action. We consider the situation where a simple
semiring acts on a semimodule. This generalizes the group
situation where $G$ is a cyclic group of prime order $p$,
i.e. where the simple ring $\Z/p\Z$ is acting on $G$ via
exponentiation. 

Simplicity of the involved semirings is important in order to
avoid Pohlig-Hellman type attacks. Using a recently found simple
semiring of order 6 we illustrate the techniques in an example.
It will require further research to assess the security of such
systems.

\def\cprime{$'$} \def\cprime{$'$}

\medskip

Paper accepted in Journal {\em Advances in Mathematics of Communications} (AMC).
\medskip
\end{document}